\documentclass[iop,apj,tighten]{emulateapj}
\usepackage{apjfonts} 
\usepackage{amsmath,amstext}
\usepackage[breaklinks,colorlinks,citecolor=blue,linkcolor=magenta]{hyperref} 
\usepackage[all]{hypcap} 
\usepackage{threeparttable}
\usepackage{bm}
\usepackage{ulem}
\usepackage{longtable}

\newcommand{\kms}{km\,$\rm s^{-1}$}
\newcommand{\angst} {\mbox{$\:${\rm \AA}}}

\newcommand{\sqdeg} {${\rm deg}^2$ }

\shorttitle{The metallicity and systemic radial velocity of RRLs}
\shortauthors{Liu, et al.}
\begin{document}
\title{Probing the galactic halo with RR Lyrae stars I: The catalog}
\author{
G.-C. Liu\altaffilmark{1}, 
Y. Huang\altaffilmark{2,9},
 H.-W. Zhang\altaffilmark{3,4},
M.-S. Xiang\altaffilmark{5},
J.-J., Ren\altaffilmark{6},
B.-Q. Chen\altaffilmark{2},
H.-B. Yuan \altaffilmark{7},
C. Wang\altaffilmark{3},
Y. Yang \altaffilmark{2},
Z.-J. Tian \altaffilmark{8},
F. Wang \altaffilmark{3,4},
X.-W. Liu\altaffilmark{2,9}
}

\altaffiltext{1}{China Three Gorges University, Yichang 443002, P.\,R.\,China.}
\altaffiltext{2}{South-Western Institute For Astronomy Research, Yunnan University, Kunming 650500, P.\,R.\,China.}
\altaffiltext{3}{Department of Astronomy, Peking University, Beijing 100871, P.\,R.\,China.}
\altaffiltext{4}{Kavli Institute for Astronomy and Astrophysics, Peking University, Beijing 100871, P.\,R.\,China. }
\altaffiltext{5}{Max-Planck Institute for Astronomy, K{\"o}nigstuhl, D-69117, Heidelberg, Germany. }
\altaffiltext{6}{National Astronomical Observatories, Chinese Academy of Science, Beijing 100012, P.\,R.\,China.}
\altaffiltext{7}{Department of Astronomy, Beijing Normal University, Beijing 100875, P.\,R.\,China.}
\altaffiltext{8}{Department of Astronomy, Yunnan University, Kunming 650500, P.\,R.\,China.}
\altaffiltext{9}{Corresponding authors (YH: yanghuang@ynu.edu.cn; XWL: x.liu@pku.edu.cn).}

\begin{abstract}

We present a catalog of  5,290 RR Lyrae stars (RRLs) with metallicities estimated from spectra of the 
LAMOST Experiment for Galactic Understanding and Exploration (LEGUE) and the Sloan Extension for Galactic Understanding and Exploration (SEGUE) surveys. 
Nearly 70 per cent of them (3,642 objects) also have systemic radial velocities measured. Given the pulsating nature of RRLs, metallicity estimates are based on spectra of individual exposures, by matching them with the synthetic templates.
The systemic radial velocities are measured by fitting the observed velocity as a function of phase assuming an 
 empirical pulsating velocity template curve.
Various tests show that our analyses yield metallicities with a typical precision of 0.20\,dex and 
systemic radial velocities with uncertainties ranging from 5 to 21\,km\,s$^{-1}$ (depending on the number of radial velocity measurements available for a given star).
Based on the well calibrated near-infrared $PM_{W1}Z$ or $PM_{K_{\rm s}}Z$, and $M_{V}$-[Fe/H] relations, precise distances are derived for these RRLs.
Finally, we include Gaia DR2 proper motions in our catalog. The catalog should be very useful for various Galactic studies, especially of the Galactic halo.

 \end{abstract}

\section{Introduction} \label{sect:intro}
Probing the Galactic structure is  an important task to help 
understand the assemblage history of our Milky Way as well as that of galaxies in general. 
The Millky Way halo contains some of the oldest stars and structures found in our Galaxy, 
and  thus provides information of the earliest stage of evolution of our Galaxy. Despite its crucial importance, 
our knowledge of the stellar halo is still far from complete, partly due to the lack of large 
samples of halo tracers to probe its properties.  

Hitherto, the main halo tracers employed include
blue horizontal branch (BHB) stars \citep[e.g.][]{Xue2008, Deason2011, Deason2014, Das2016}, 
K giants \citep[e.g.][]{Xue2015, Xu2018}, near-main-sequence turnoff stars (nMSTO) \citep[e.g.][]{ Sesar2011, Pila-Diez2015}
and RR Lyrae stars (\citep[RRLs; e.g.][]{Watkins2009}. 
The available number of BHB stars is small since they are difficult to identify; distance estimation of K giants is not easy considering that 
their intrinsic luminosities vary by two orders of magnitude (depending on stellar age and metallicity);
and the luminosities of nMSTO are not high enough to be useful for probing the distant outer  halo of the Galaxy.

Compared to the other tracers, RRLs are ideal for studying the halo properties. 
First, RRLs are old ($> $9 Gyr), low-mass metal-poor
stars that reside in the instability strip of the horizontal
branch (HB) and thus represent a fair sample of  the halo
populations \citep{Smith2004}. Secondly, their well defined period-luminosity relation
makes them good standard candles, allowing one to accurately map out the 3D structure of the halo. 
Thirdly, they retain a record of the chemical composition of the environment in which they were born,
thus can be used to study the early stage chemical evolution of the Galaxy formation. 
Finally, they are relatively easy to identify based on their colors and variabilities, enabling
the construction of samples of few contaminations.
In short, RRLs are excellent tracers to study the structure, formation and evolution of the Galactic halo. 

Nevertheless, time-domain photometric surveys alone can not provide precise measurements 
of metallicity and systemic radial velocity of RRLs. Such information has to be extracted from spectroscopic observations. 
Fortunately, a number of large-scale spectroscopic surveys have been carried out in the past two decades, 
including the RAVE  \citep{Steinmetz2006}, the SDSS/SEGUE  \citep{Yanny2009}, 
the SDSS/APOGEE \citep{Majewski2017}, the LAMOST \citep{Deng2012, Liu2014} and the GALAH \citep{DeSilva2015} surveys.

The spectroscopic data, combined with information (e.g. period, phase and amplitude) derived from light curves
provided by photometric surveys and astrometric information (e.g. parallax and proper motions) from the Gaia DR2, 
allow one to construct a large sample of RRLs with full phase space information of 3 dimensional  position and velocity, as well as of metallicity, 
and to use the sample to probe the formation and evolution of the Galactic halo.

Values of metallicity and radial velocity of RRLs can not be measured by treating them as normal, steady stars, 
as RRLs are pulsating and their spectra vary with time on short time scales.
To measure the metallicities of RRLs, the most precise method is to utilize 
high resolution spectra. High resolution spectroscopy is however quite costly of big telescope time,
and no more than a hundred bright local RRLs have been observed this way \citep[e.g.][]{For2011, Kinman2012, Nemec2013, Govea2014, Pancino2015}.
For low resolution spectra, the traditional method to measure the metallicities of RRLs
 is  the so-called $\Delta S$ method, first proposed by \citet{Preston1959}. The $\Delta S$ index describes
 the difference between the spectral types as given by the H~{\sc i} Balmer lines and by the Ca~{\sc ii} K line measured with low-resolution spectra obtained at minimum light.
Many studies have been carried out with this method to derive metallicities of RRLs, 
by calibrating the $\Delta S$ relation with either spectroscopic or photometric observations \citep[e.g.][]{Butler1975, Freeman1975, Layden1994}. Although this method is efficient, 
the scale relation between $\Delta S$ and [Fe/H] show some nonlinear variations in some cases, 
leading to some random and systemic errors in the final results.
 
 In this paper, we present a new method to measure the metallicities of RRLs. 
 For the modern large-scale spectroscopic surveys (e.g. the LAMOST and  SDSS surveys),  
 there are often two to three exposures made per visit, yielding two to three spectra on a short time span. 
 In addition to the multiple spectra from a single visit, some fields are visited more than once.
Excluding spectra affected by shock waves, the metallicity can be measured from the individual single-exposure 
 spectra with a template matching technique (see Section\,3.1 for detail).
 The mean value, weighted by the signal-to-noise ratios (SNRs), of estimates yielded by the individual single-exposure spectra,  is then adopted as the final metallicity of  the RRL.
 
  In addition to metallicity, it is also crucial to precisely measure the systemic radial velocities of RRLs since they are of vital importance for a variety of  Galactic studies, such as identifying halo substructures and exploring their origins by
kinematics  \citep[e.g.][]{Vivas2001, Keller2008, Miceli2008, Casetti2009, Watkins2009, Carlin2012}, and constraining the mass distribution of the Milky Way \citep[e.g.][]{Xue2008, Huang2016}.

However, accurately measuring systemic velocities of RRLs is quite challenging since the observed radial velocities contain a pulsation component which, of a typical amplitude of several tens of km\,s$^{-1}$, needs to be accounted for.
To derive the values of systemic velocity (${\rm RV}$), two approaches are generally adopted.
In one approach, one schedules the observations at the right phase (i.e. $\sim 0.5 \pm 0.1$) such that the pulsation has nearly zero contribution to the observed radial velocity, i.e. $\rm{RV}_{\gamma} = \rm{RV}_{\rm obs}$. 
This approach is however not suitable for data collected with large scale, multi-object spectroscopic surveys as in our case. Alternatively, one can correct the measured radial velocities for contribution of pulsation assuming a pulsation model (or an empirical template), that describes the pulsation  velocity as a function of phase. In this paper, we utilize the latter approach to measure the systemic velocities of RRLs, by adopting the empirical template of radial velocity curves of {\it ab}-type RRLs as constructed by \citet{Sesar2012}.

 The paper is the first one in a series that utilize RRLs to explore the formation and evolution of the Galactic halo. 
 The data used in the current work is described in Section\,2. Estimation of metallicities of RRLs from the spectra is introduced in Section\,3.
 Determinations of systemic radial velocities and distances are presented in Sections\,4 and 5, respectively.
 In Section\,6, we present the final catalog and describe its general properties. Finally, a summary is given in Section\,7.
 
\section{Data}\label{sect:data}

In this Section, we first collect known RRLs identified in the various time-domain photometric surveys or variable source catalogs. 
Then available spectra are searched for those stars in the various large-scale spectroscopic surveys.

\subsection{RR Lyrae stars from photometric surveys}

To collect known RRLs, we use the catalogs of variable stars  from the QUEST \citep{Vivas2004, Mateu2012, Zinn2014},
 NSVS \citep{Kinemuchi2006, Hoffman2009}, LONEOS \citep{Miceli2008},  
 GCVS \citep{Samus2009}, LINEAR \citep{Sesar2013} and the Catalina  \citep{Drake2013, Drake2014} surveys. 
Data of the southern Hemisphere are not included since only spectroscopic data in the northern sky are used (see next Section).
A prior is set to each survey (see Table\,1), according to their observational epochs (higher prior for those surveys more close to the spectroscopic observations described as follows) and the typical number of photometric observations.
 For each survey,  we compile all the available parameters of identified RRLs  
into a  single catalog, including period, amplitude, epoch, mean $V$-band magnitude, variable star type and distance if derived.
 In total, we obtain a list of 32,243 unique RRLs from those surveys.

\begin{figure*}
\centering
\includegraphics[width=16cm]{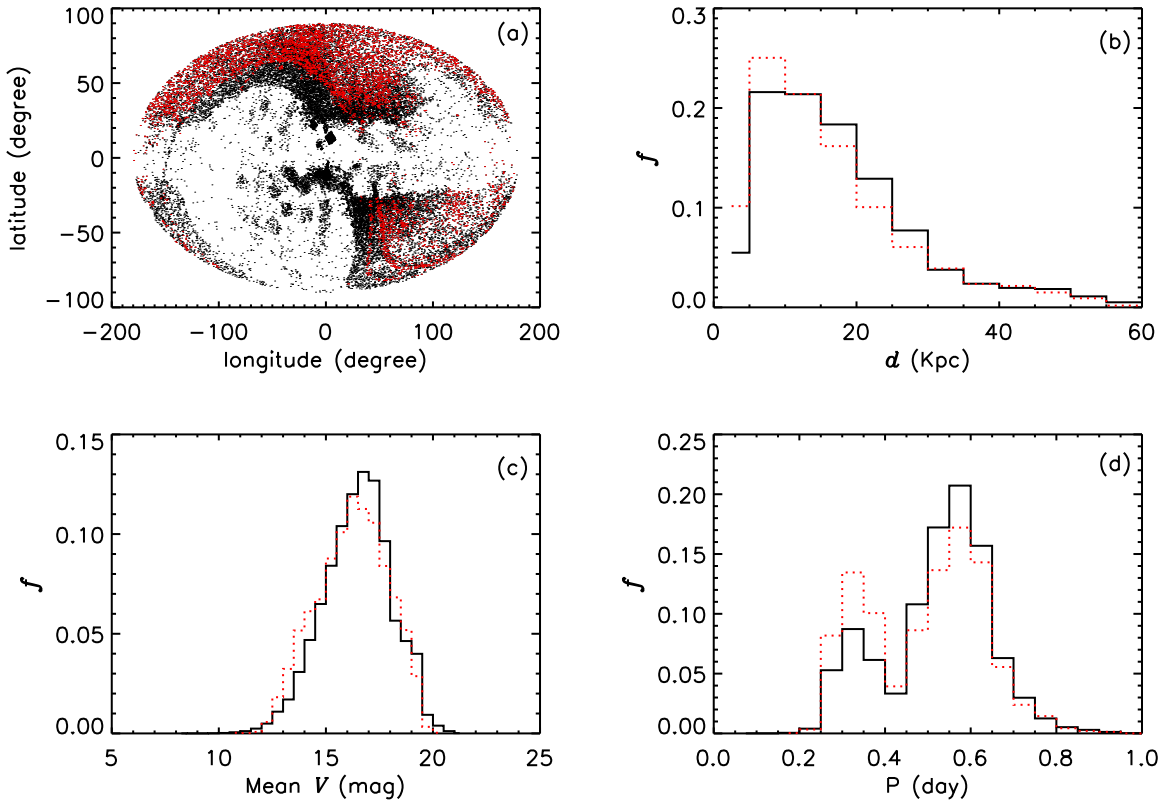}
\caption{ Basic properties of our photometric (black dots/lines) and spectroscopic (red dots/lines) RRL samples. Panel (a) shows the spatial distribution in Galactic coordinates, Panel (b) the normalized distribution of distances collected from the literature, Panel (c) the normalized distribution of mean $V$ band magnitudes and Panel (d) the normalized distribution of periods.} 
\label{fig:photo}
\end{figure*}

   Table\,\ref{tab:survey} summarizes the properties of the individual surveys included in the current study. Fig.\,\ref{fig:photo} plots the  distributions of the stars in Galactic coordinates, 
  in distances (adopted directly from the literature), in mean $V$ band magnitudes and in periods. If a parameter of a given star is available from more than one survey, then the value from the survey with highest prior is adopted. 
  
   Finally, we note that the catalogs of \citet{Drake2013b} and \citet{Drake2014} do not provide the epochs of maximum 
   light for the catalogued RRLs. For those stars, we have calculated the missing values by ourselves from the light curves provided by the Catalina survey\footnote{http://nesssi.cacr.caltech.edu/DataRelease/RRL.html}.

 \begin{table*}
 \begin{minipage}{160mm}
 \centering
\caption{Recent large-scale photometric surveys of RRLs}
\begin{footnotesize}
\begin{tabular}{lcccccccc}
\hline
\hline
Survey & Filters & Area (\sqdeg) & Range of V magnitude & The typical number of  & Obsevation year & Sources  & Prior & Reference \\
            &            &                        &                                     & photometric observations&                         &                &           &                 \\
\hline
Catalina      & $V$     & $\sim33,000$ & $ 12-20$ &  $60\sim 419$ &$2004-2011$&  $23,306$      &5     & 1\\

QUEST  & $UBVRI$ & $380/476$ & $13.5-19.7$   &$15\sim 40$ &$1998-2008$&   $1,857$         &2   & 2\\
             
NSVS         & ROTSE-NT & $\sim31,000$       & $V<14$ &$100\sim 500$& $1999-2000$ &  $1,304$  &0   &3\\

LINEAR     &   no spectra filter      & 8,000            & $14-17$ &  $200\sim460$ &$1998-2009$ &    $5,684$  & 4 & 4 \\

LONEOS    & LONEOS-NT & $1,430$        & $V<18$  & $28\sim50$ & $1998-2000$ & $838$          &0  &  5 \\

SDSS Str82 & $ugriz$  & $249$     & $15-21$  & $30\sim40 $ &$1998-2006$ &  $601$                     &3   &  6 \\

GCVS        &    --             &   --             &    --     &    --  &    --            &    $7,954$    &1   &7 \\                                
\hline
\end{tabular}
\begin{flushleft}
\textbf{Notes.} The references are: 1-- \citet{Drake2013, Drake2014} ; 2--\citet{Vivas2004, Mateu2012,  Zinn2014}; 3--  \citet{Kinemuchi2006, Hoffman2009} ; 4--\citet{Sesar2013};
                        5--\citet{Miceli2008}; 6--\citet{Watkins2009, Sesar2010, Suveges2012}; 7-- \citet{Samus2009}.
\end{flushleft}
\end{footnotesize}
\label{tab:survey}
\end{minipage}
\end{table*}

\subsection{Spectroscopy}

   In the current work, our major spectroscopic data set comes from the LAMOST Galactic spectroscopic surveys \citep{Deng2012, Zhao2012, Liu2014}.
    LAMOST (also named Guoshoujing Telescope) is an innovative quasi-meridian reflecting Schmidt telescope capable of simultaneously recording spectra of up to 4000 objects in a large field of view (FoV) of 5 degrees in diameter. The spectra cover the wavelength range from 3700 to 9100 \angst  \,with a resolving power $R \sim 1800$.
Typically 2 $\sim$ 3 exposures are obtained for each plate, with typical integration time per expsystemic velocityosure of 10 to 40 minutes, depending on the brightness of targeted sources. Since the LAMOST FoV is circular, field overlapping cannot be avoided in order to achieve a contiguous sky coverage. About 25 percent of all targets have been observed twice and over 2 percent  three times \citep{Yuan2015}. This greatly benefits the measurements of systemic radial velocity of RRLs reported in this work. By June 2016, the LAMOST Galactic surveys, initiated in October 2012, have obtained $\sim$ 6 million  quality spectra, mostly of Galactic stars.  This number is still increasing at a rate of ~1 million per annum.
    
 Another main source of spectra comes from the SDSS/SEGUE \citep{Yanny2009}.
As a major component of the SDSS-II, SEGUE operated from 2005 August to 2008 July, and obtained more than 240,000 spectra of Galactic  stars of magnitudes $14.0 < g <20.3$, with a  spectral coverage and resolution similar to those of LAMOST. In order to obtain spectra of sufficient signal-to-noise ratios (SNRs), the typical total integration time for bright plates of sources of ($14.0 < r < 17.8$) is 1 hr and that for plates of fainter sources of $17.8 < r <20.1$ is 2 hr. The integration time of the individual exposures ranges between 10 $-$ 30 minutes.  SEGUE-2, the successor of SEGUE, obtained additional 155,520 spectra with the same instrument. All the data from SEGUE and SEGUE-II are included in the SDSS Data Release 12 \citep[SDSS DR12;][]{Alam2015}.
    
  The spectral database of SDSS DR12 is downloaded and then cross-matched with the aforementioned compiled catalog of photometrically identified RRLs.
  In total, 3,834 common stars are found, with a total of 20,772 single-exposure spectra. Similarly, a total of 3,016 common sources 
  (with a total of 10,667 single-exposure spectra) are found between the photometric catalog of RRLs and the LAMOST DR2 of value-added catalog \citep{Xiang2015, Xiang2017}.
  By combining the two data sets, a total of 6,268 RRLs with a total 31,439 single-exposure spectra are obtained. Typical SNRs of those single-exposure spectra are  around 15.

\section{spectroscopic metallicities}

   As mentioned, we utilize single-exposure spectra instead of those combined to measure the 
metallicities, given the pulsating nature of RRLs.

\subsection{Measurement method}

    To estimate the metallicities of RRLs from single-exposure spectra, we adopt a template matching method by comparing the observed spectra with the synthetic ones based on the least-$\chi^2$ technique. 
    The synthetic spectra library was generated with code SPECTRUM \citep{Gray1999} of version 2.76, 
    utilizing the Kurucz stellar model atmospheres of \citet{Castelli2004} that cover the wavelength range from 3850 to 5600 $\angst$ at a resolution of 2.5  \angst.
    We degrade the model spectral resolution to match that of  LAMOST and SDSS (R $\sim$ 1800). 
     Considering the typical ranges of atmospheric parameters of RRLs, we limit our synthetic spectra to parameter ranges: 
     effective temperature 6000 $ \leq $ $T_{\rm eff}$ $ \leq $ 7500 K  in step of 100 K, surface gravity 1.5 $ \leq $ log\,$g$ $ \leq $ 4.0 \,dex
    in step of 0.25 dex, and metallicity $-$3.0 $ \leq $ [Fe/H] $ \leq $ 0.0 \,dex in step of 0.1 dex. Considering the old (typically > 9 Gyr) and metal-poor nature of RRLs, we fix the value of $\alpha$-element to iron abundance ratio $[\alpha/ \rm Fe]$  to 0.4.
    
\begin{figure*}
  \centering
  \includegraphics[width=16 cm, angle=0]{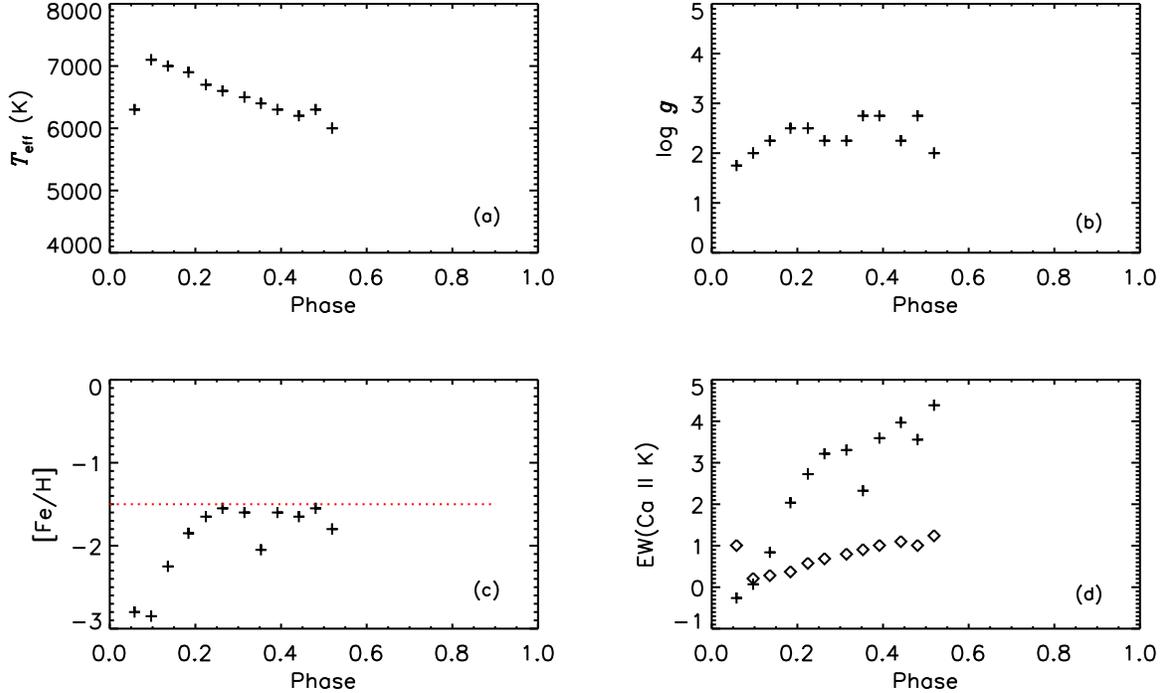}
  \caption{The atmospheric parameters of single-exposure spectra obtained 
    by template matching for SDSS J134134.54+281855.2 with twelve times exposures, which is a member star of globular cluster NGC 5272. Panel (a) shows the effective temperature varies with phase, Panel (b) the log\,$g$ varies with phase, Panel (c) the measurement of metallicity at different phases, the dash line shows the reference metallicity given by \citet{Harris2010}, Panel (d) the equivalent width of Ca {\sc ii} K line EW(Ca~{\sc ii} K) at different phases, respectively.  In the Panel (d), the plus symbols represent the calculated values of EW(Ca~{\sc ii} K) from twelve single-exposure spectra and the diamond symbols indicate the minimal value of EW(Ca~{\sc ii} K) of model spectra with the same $T_{\rm eff}$ as that from observed single-exposure spectra (estimated by the first round of fitting, see Section 3.1 for details).
  }
   \label{parameter}
\end{figure*}

    In order to more precisely obtain the parameters $T_{\rm eff}$ and  [Fe/H], we match the observed spectra with the synthetic ones by two steps. Firstly, we measure the effective temperature $T_{\rm eff}$ by least-$\chi^2$ fitting. Considering that the Balmer lines are most sensitive to effective temperature, we give twice weights (weighted by the inverse variances of the spectral fluxes) to spectral pixels that cover for ${\rm H\alpha}$, ${\rm H\beta}$, and ${\rm H\gamma}$ lines when comparing the full observed spectra with the synthetic ones of wavelength range from 3850 to 5600 $\angst$ pixel by pixel. At this step, effective temperatures are well determined from single exposure spectra.\footnote{Surface gravity and metallicity are also estimated at this step but they are not well constrained due to high weights on Balmer lines and thus not been used.} The derived effective temperatures are then set as input values at the following second round of fitting. In the second step, values of the Ca~{\sc ii} K line (mostly sensitive to metallicity) and the continum with the synthetic one but fixing $T_{\rm eff}$ to the value deduced in the first step. In this step, spectral pixels covering 
 Ca{\sc ii} H and the Balmer lines, i.e. pixels of the wavelength ranges 3960 to 3980, 4092 to 4112, 4330 to 4350 and 4851 to 4925\angst\, are masked out. Best-fit values of [Fe/H] and  log\,$g$ yielded this second round of optimization are adopted for the star. 
   Fig.\,\ref{parameter} plots the resulted stellar atmospheric parameters of an RRL as a function of phase. For this particular target, a total of  12 single-exposure spectra are available. We find that the estimated values of $T_{\rm eff}$ and  log\,$g$ vary with the phase of pulsation, largely in  consistent with the theoretical predictions.

    By above two steps, the metallicities of RRLs are obtained from the individual single-exposure spectra. 
   In principle, no matter what the pulsation stage of the RRL is when targeted, its metallicity should unchange and keep the same value. 
   However, due to the effects of shock waves on the hydrogen and metal lines of RRL spectra \citep{ Fokin1992, Gillet2004, Pancino2015}, the estimated [Fe/H] could change significantly depending on the phase when the spectrum was taken.
   As  Fig.\,\ref{parameter} shows, the metallicities of an RRLs estimated during phase between 0 and 0.15 vary dramatically, 
     reflecting the significant effects of shock waves during those phases during the pulsation cycle.
      To avoid the potential bias in the metallicity determinations caused by shock waves, we try to 
      exclude single-exposure spectra possibly affected by the shock effects for parameter estimates. 
      In general, shock waves mostly occur during the phase 0 to 0.15, and 0.85 to 1.0,
      but they also can happen at other phases. Considering that the Ca~{\sc ii}  K line is  easily affected by the shock wave effects, here, we adopt its equivalent width (EW)
      as a criteria to asses whether a spectrum in concern is affected by shock waves or not.

   We first calculate the values of equivalent width of the Ca~{\sc ii} K line, EW(Ca~{\sc ii} K), of all model spectra and find the minima for each effective temperature.
Then we fit the minima as a function of temperature with a second-order polynomial and find Min[EW(Ca~{\sc ii} K)]=15.62\,$-$\,0.0037\,$T_{\rm eff}$\,+\,2.21$\times10^{-7} T_{\rm eff}^2$.
  At the same time, we also calculate the EW(Ca~{\sc ii} K) for each of the single-exposure spectra of our sample stars. As Fig.\,\ref{parameter} shows, we find that the values of metallicity measured from single-exposure spectra of EW(Ca~{\sc ii} K) less than the corresponding Min[EW(Ca~{\sc ii} K)], are different from those measured at other phases [c.f. Panel (d) of Fig.\,\ref{parameter}]. This indicates that those former spectra are affected by shock waves and metallicities yielded by those spectra are consequently ignored. The final adopted metallicity and its error are weighted mean by its errors, yielded by single-exposure spectra (see the Section 3.2.1 ), if having metallicity measurement greater than 2. If only one single spectrum available for a star, we directly use the metallicity of this single spectrum as the final adopted value of this star. Finally, we obtain the metallicities for 5,290 RRLs. It should be noted that, \citet{Fabrizio2019} also have estimated the metallicities of 2,382 fundamental RRLs by $\Delta S$ method very recently.
      
      By comparing the metallicity estimate of  common stars selected from LAMOST and SEGUE surveys with similar SNRs (i.e., $\Delta$\, SNR\,$\leq 5$), a negligible offset (around 0.04\,dex) is found between the metallicity measurements from the spectra obtained by the two surveys. We therefore assume the metallicitiy scales yielded from the two surveys are the same.

\subsection{Validation of metallicities}

    In this Section, we examine the accuracy of metallicities of RRLs measured by above method in the following ways: 
    1) Check the internal uncertainties using duplicate-observations; 2) Check both the random and systemic errors by 
    comparing with metallicity measurements from high resolution spectroscopy. 
    
\subsubsection{Comparison of results from multi-epoch observations}

\begin{figure*}
  \centering
  \includegraphics[scale=0.6, angle=0]{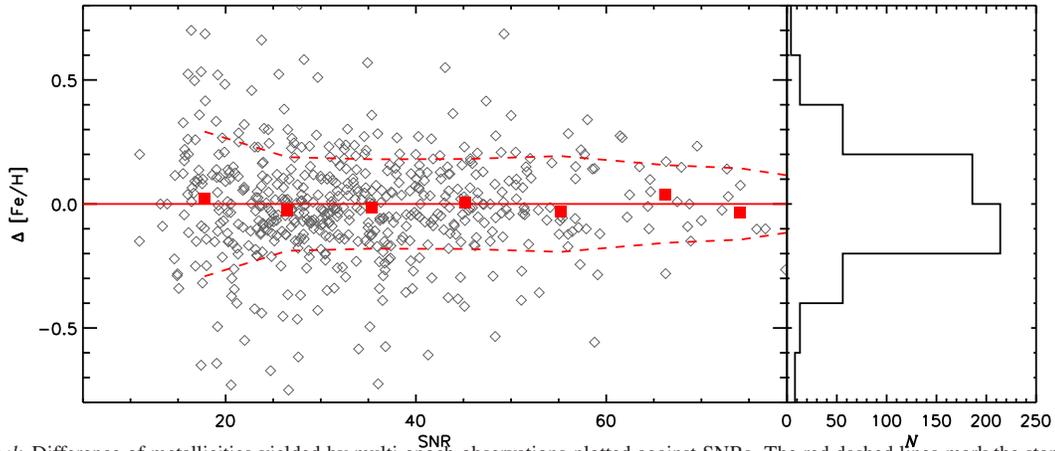}
  \caption{ $Left panel$: Difference of metallicities yielded by multi-epoch observations plotted against SNRs. The red dashed lines mark the standard deviations. The red squares indicate the average of the differences in the individual SNR bins and the red solid line delineates zero differences.
    $Right panel$: Histogram of the differences.
          }
   \label{fig:multiobs}
\end{figure*}

To estimate the internal errors of the metallicities derived, we use multi-epoch observations of our sample stars.
Doing so, the differences of two metallicity measurements of similar SNRs (i.e. $\Delta \rm{SNR} < 10$) as a function of the mean SNR are shown in Fig.\,\ref{fig:multiobs}. 
As the figure shows, the median differences are almost zero, with no significant systemic trend.
As expected, the standard deviations of the differences decreases with SNR.
We fit the standard deviations (divided by $\sqrt{2}$) as a function of SNR, and find  ${\rm s.d./\sqrt{2}}=0.08+2.04/\rm{SNR}$. For the observations reported here, the typical standard deviation is about 0.2 dex. 
   We use the standard deviations derived by above function as the error ($\sigma_{i}$) of the metallicity estimated by individual single-exposure spectrum when it's SNR less than 40, 
   and for SNR $\geq$40, the error are fixed to the value of $0.08+2.04/40=0.13$  dex.

 \subsubsection{Comparison of results with reference stars}
  
 In order to check the zero points of our metallicity measurements, we collect reference stars from the literature with reliable metallicity estimates
 either from high resolution spectroscopy or as a member star of a globular cluster(GC).
For GC member stars, we use a compiled catalog of over 3000 variable stars in 103 GCs \citep{Clement2017}.
The properties (e.g. metallicities, radial velocities and distances) of those GCs are taken from \citet{Harris2010}. 
The metallicity scale adopted here is the one established by \citet{Carretta2009}, which is a fundamental shift from the older metallicity scale \citep{Zinn1984} with superior abundance analysis methods based on more advanced model atmospheres.
In addition, we have collected stars with metallicity and systemic radial velocity estimates measured with high resolution spectroscopy
 \citep[e.g.][]{Clementini1995, For2011, Kinman2012, Nemec2013, Govea2014, Pancino2015}.
 We have cross-matched  our RRLs spectroscopic sample with above compiled catalogs and obtained 47 stars in common.
 Those common stars form our reference star sample. Tables\,\ref{tab:ref_star1} and \ref{tab:ref_star2}  present respectively relevant information 
 of the reference stars, for GC members and from high resolution spectroscopy.
  
 \begin{table*}
 \begin{minipage}{160mm}
 \centering
\caption{Parameters of reference stars from globular clusters}
\begin{footnotesize}
\begin{tabular}{lccccccc}
\hline
\hline
Cluster & RA\,(degree) & Dec\,(degree) & [Fe/H]  & RV\,(\kms)  & $\rm{RV}_{err}$\,(\kms) & Distance\,(kpc) & $N$  \\
\hline
NGC 4147&       182.544&       18.581&     $-$1.80&      183.2&     0.70&      19.3&           1\\
NGC 5053&       198.997&       17.741&     $-$2.27&      44.0&     0.40&      17.4&           2\\
NGC 5024&       198.359&       18.162&     $-$2.10&     $-$62.9&      0.30&      17.9&           3\\
NGC 5272&       205.392&       28.507&     $-$1.50&     $-$147.6&      0.20&      10.2&          11\\
NGC 5466&       211.373&       28.507&     $-$1.98&     $-$106.9&      0.20&      16.0&           2\\
NGC 5904&       229.827&       2.283&     $-$1.29&      53.2&       0.40&      7.5&           1\\
NGC 6341&       259.300&       43.207&     $-$2.31&     $-$120.0&       0.10&      8.3&           1\\
NGC 7089&       323.471&     $-$0.799&     $-$1.65&     $-$5.3&        2.0&      11.5&           2\\
NGC 7078&       322.580&       12.316&     $-$2.37&     $-$107.0&       0.20&      10.4&           2\\                                                                                      
Pal 5         &       228.991&     $-$0.190&     $-$1.41&     $-$58.7&       0.20&      23.2&           4\\
\hline
\end{tabular}
\end{footnotesize}
\label{tab:ref_star1}
\end{minipage}
\end{table*}
 
\begin{table*}
 \begin{minipage}{160mm}
 \centering
\caption{Metallicity of reference stars from high-resolution spectroscopy}
\begin{footnotesize}
\begin{tabular}{lccccc}
\hline
\hline
 Star\,&RA\,(degree) & Dec\,(degree) & \rm{[Fe/H]} & $\rm{[Fe/H]}_{err}$ & Reference \\
\hline
       DR And&16.295&       34.218&     $-$1.37&  0.12&   Pancino et al.(2015)\\
       BK Eri&42.483&      $-$1.420&     $-$1.72&    0.21&   Pancino et al.(2015)\\
       SZ Gem&118.431&       19.273&     $-$1.65&   0.07& Pancino et al.(2015)\\
       SS Leo&173.477&    $-$0.033&     $-$1.48&    0.07&  Pancino et al.(2015)\\
       UV Vir&185.320&      0.368&     $-$1.10&     0.12&  Pancino et al.(2015)\\
       UZ CVn&187.615&       40.509&     $-$2.21&  0.13&  Pancino et al.(2015)\\
       RV UMa&203.325&       53.988&     $-$1.20&  0.08&  Pancino et al.(2015)\\
       TW Boo&221.275&       41.029&     $-$1.47&   0.05&  Pancino et al.(2015)\\
       VIII-14&256.891&       58.850&     $-$2.92 & - &        Kinman et al. (2012)\\
       V355 Lyr&283.358&       43.155&     $-$1.14&  0.17&    Nemec et al. (2013)\\
        KIC 11125706&285.245&       48.745&     $-$1.09&   0.08&   Nemec et al. (2013)\\
        NQ Lyr&286.952&       42.300&     $-$1.89&   0.10&  Nemec et al. (2013)\\
       NR Lyr&287.114&       38.813&     $-$2.54&   0.11&   Nemec et al. (2013)\\
        FN Lyr&287.593&       42.459&     $-$1.98&   0.09&  Nemec et al. (2013)\\
        V838 Cyg&288.516&       48.200&     $-$1.01&   0.10&   Nemec et al. (2013)\\
        V1104 Cyg&289.502&       50.755&     $-$1.23&  0.15&    Nemec et al. (2013)\\
        V1107 Cyg&289.939&     47.1012&    $-$1.29&   0.23&  Nemec et al. (2013)\\
        V2470 Cyg&289.991&       46.889&    $-$0.59&   0.13&   Nemec et al. (2013)\\
       V894 Cyg&293.254&       46.240&     $-$1.66&  0.12&   Nemec et al. (2013)\\

\hline
\end{tabular}
\end{footnotesize}
\label{tab:ref_star2}
\end{minipage}
\end{table*}

\begin{figure*}
  \centering
  \includegraphics[scale=0.6, angle=0]{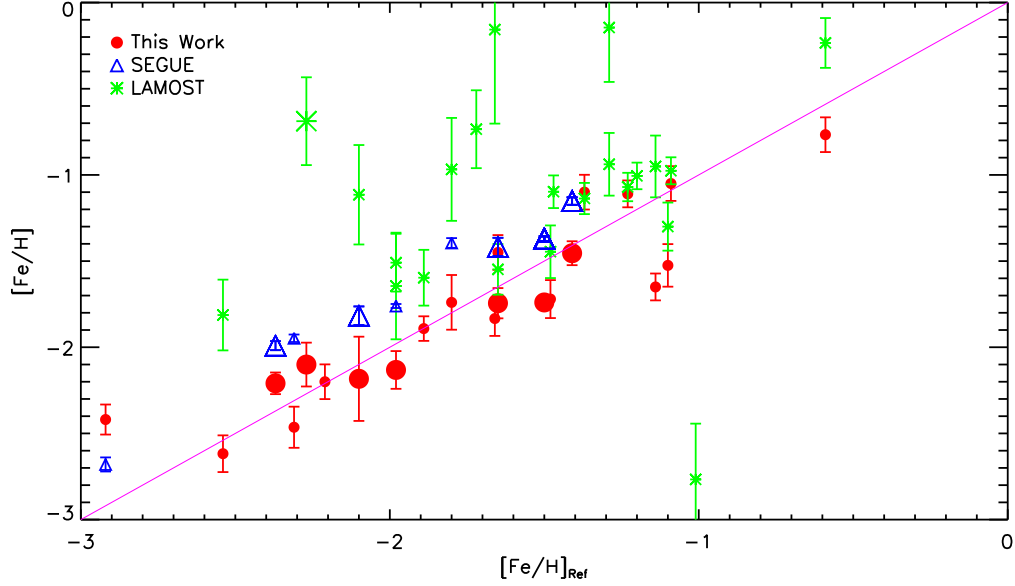}
  \caption{Metallicities of reference stars estimated in the current work (red dots), and those given by the default SEGUE (blue triangles ) 
  and LAMOST (green stars) pipelines, are plotted against the literature values. Those stars in globular clusters are averaged in a single dot and represented by larger symbols.}
   \label{fig:segue}
\end{figure*}

As Fig.\,\ref{fig:segue} shows, the values of metallicity estimated in the current work 
 agree well with those of the compiled reference stars, with a negligible offset ($-$0.04) and a standard deviation of 0.22\,dex.
 The dispersion is comparable to that yielded by multi-epoch observations. 
  All the tests manifest that intrinsic consistency of our measurements is good.  
  
 In addition, we have also compared the metallicity measurements yielded by the default SEGUE and LAMOST pipelines with the literature values for reference stars.
  As Fig.\, \ref{fig:segue} shows, both pipelines over estimate the metallicities, sigificantly.

\section{systemic radial velocities}

As mentioned in Section \ref{sect:intro}, it is of crucial importance to measure the systemic radial velocities of RRLs. In this Section,  we use the empirical template radial velocity curves of  RRLs to fit the observed radial velocities as a function of phase as derived from the individual  single-exposure spectra of RRLs in order to obtain their systemic radial velocities. 

\subsection {Measurement method}
\label{RV method}
Here, we adopt the empirical template radial velocity curves of {\it ab}-type RRLs constructed by \citet{Sesar2012} for H$\alpha$, H$\beta$ and H$\gamma$ lines. According to Fig.\,3 of \citet{Sesar2012}, the uncertainty of systemic velocity yielded  by fitting the empirical templates  increases dramatically 
for observational phase  greater than 0.7. On the other hand, spectra at phases less than 0.1 are liable to strong effects of shock waves. Consequently, we decide to only use single-exposure spectra taken at  phases 
between 0.1 and 0.7. Amongst those, spectra affected by shock waves , as implied by the criteria  EW(Ca~{\sc ii} K) < Min[EW(Ca~{\sc ii} K)], are further excluded when deriving  the systemic velocity by fitting with the radial velocity template. 

\begin{figure*}
  \centering
  \includegraphics[scale=0.6, angle=0]{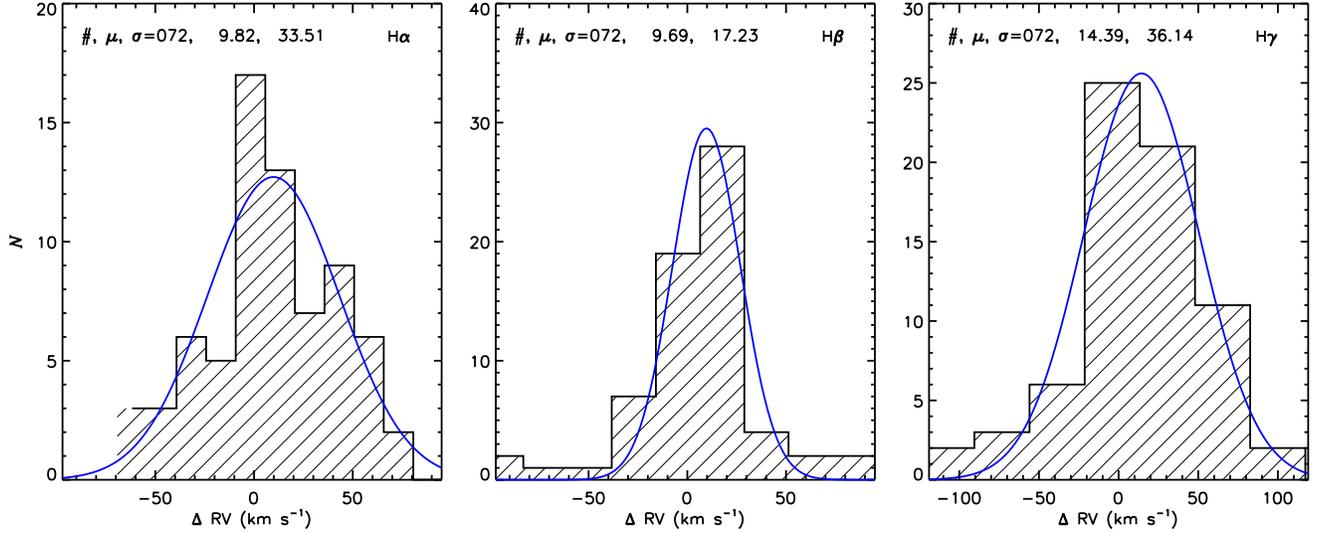}
  \caption{ Distributions of values of ($\rm{RV}_{LAMOST} - \rm{RV}_{SDSS}$ ) as measured from ${\rm H\alpha}$ (left),
${\rm H\beta}$ (middle), and from ${\rm H\gamma}$ (right) lines, respectively. The three numbers marked in the top left of each panel denote the number of common stars, the mean and the standard deviation of the RV difference between LAMOST and SDSS, respectively. Blue lines represent the Gaussian fit.}
   \label{fig5}
\end{figure*}

For those adopted single-exposure spectrum, we derive the observed RV by fitting ${\rm H\alpha}$, ${\rm H\beta}$, and ${\rm H\gamma}$ line profiles with a Gaussian function, together with a first-order polynomial, and measure their centers (wavelength coverage from 4325 to 4357 \angst\, for ${\rm H\gamma}$ line, 4845 to 4878\angst\, for ${\rm H\beta}$ line and 6548 to 6580 \angst\, for ${\rm H\alpha}$ line, are respectively used in the fits.).
Considering that we use spectra from two surveys (LAMOST and SDSS), radial velocities derived from spectra of the two surveys need be calibrated to a common scale.
Doing so, common sources from the two surveys of similar phases ($\Delta \rm{phase} < 0.05$) and high SNRs (> 50) are selected, yielding a total of 72 targets. 
Distributions of the RV differences (LAMOST values minus SDSS ones) for different lines (i.e. ${\rm H\alpha}$, ${\rm H\beta}$, and ${\rm H\gamma}$)  are shown in Fig.\,\ref{fig5}.
The medians and standard deviations estimated from the distributions are, respectively, $9.82 \pm 33.51$, $9.69 \pm 17.23 $, and  $14.39 \pm 36.14$  \kms for 
${\rm H\alpha}$, ${\rm H\beta}$, and ${\rm H\gamma}$. We then calibrate LAMOST RVs derived from the three lines to the scales of SDSS, using the median differences found above.

\begin{figure*}
  \centering
  \includegraphics[scale=0.65, angle=0]{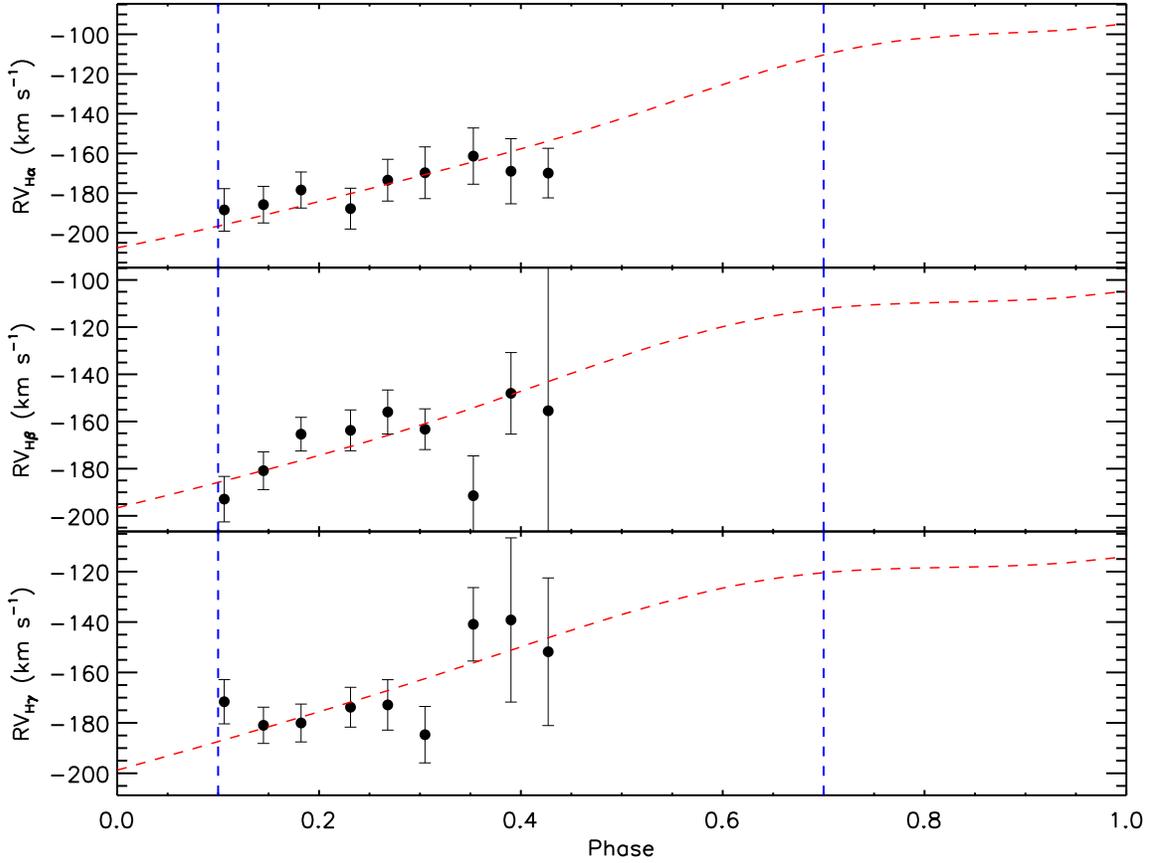}
  \caption{ As an example,  radial velocities (black dots with error bars)  measured from  ${\rm H\alpha}$ (top panel),
${\rm H\beta}$ (middle panel), and from ${\rm H\gamma}$  (bottom panel) are fitted with the radial velocity  templates constructed by \citet{Sesar2012} (red dashed line), 
 the blue dashed lines marked the fitting range (0.1 $ \leq$ phase $\leq$ 0.7).
          }
   \label{fig:fitrv}
\end{figure*}

For a given star that has  $\rm{RV}_{obs}$ measured at several phases, one can derive the systemic velocity $\rm{RV}$ by fitting the following equation,

\begin{equation}
\label{eq}
\rm{RV}_{obs}\,(\Phi_{obs}) = A_{rv}T\,(\Phi_{obs}) + \rm{RV},
\end{equation}
where ${\rm \Phi_{obs}}$ is the observational  phase, ${\rm A_{rv}}$ the amplitude of the radial velocity curve, fixed to the
 mean values reported in Table 1. of \citet{Sesar2012}, i.e. 111.9,  90.9  and 82.1  \kms  for 
 ${\rm H\alpha}$, ${\rm H\beta}$, and ${\rm H\gamma}$ respectively,
 and ${\rm T\,(\Phi_{obs})}$ the radial velocity curve template. If only one radial velocity measurement is available, we 
 directly interpolate the template to get the systemic radial velocity. 
 As an example, Fig.\,\ref{fig:fitrv} shows  the fit for a  source with 9 phase data points.

With above procedure, the final systemic radial velocity adopted for a given star is given by the weighted mean (by uncertainties) of measurements yielded by the three Balmer lines, i.e. $\rm{RV}_{ H\alpha}$, $\rm{RV}_{ H\beta}$  and $\rm{RV}_{ H\gamma}$.
 Basically,  the measurement uncertainties are highly dependent on the number of radial velocity measurements used in the fitting.
 
 To derive the random errors of systemic velocity measurements from the individual lines, we solve the following equations:
 
 \begin{equation}
\sigma_{\alpha}^{2} (n) + \sigma_{\beta}^{2} (n) = \sigma_{\alpha\beta}^{2} (n)\text{,}
\end{equation}
\begin{equation}
\sigma_{\alpha}^{2} (n)+ \sigma_{\gamma}^{2} (n) = \sigma_{\alpha\gamma}^{2} (n)\text{,}
\end{equation}
\begin{equation}
\sigma_{\beta}^{2} (n) + \sigma_{\gamma}^{2} (n) = \sigma_{\beta\gamma}^{2} (n) \text{,} 
\end{equation}  

Here $n$ is the number of available radial velocity measurements for a given star and $\sigma_{\alpha\beta}(n), \sigma_{\alpha\gamma}(n)$, and $\sigma_{\beta\gamma}(n)$ 
are standard deviations of the  differences of systemic velocities measured from two out of three lines for given $n$.

Table \ref{tab:error} lists those $\sigma$ values for different values of $n$ for  ${\rm H\alpha}$, ${\rm H\beta}$, and ${\rm H\gamma}$ lines. 
We compare values of  $\rm{RV}_{ H\alpha}$ and $\rm{RV}_{ H\beta}$  with those of $\rm{RV}_{ H\gamma}$ and obtain mean values of differences 14.59, and 5.83 \kms  for $\rm{RV}_{ H\alpha} - \rm{RV}_{ H\gamma}$, and $\rm{RV}_{ H\beta} - \rm{RV}_{ H\gamma}$, respectively. The non-zero mean RV differences between H$\alpha$, H$\beta$ and H$\gamma$ lines are largely due to the existing of a wavelength-dependent residual after the wavelength calibration.  We therefore convert $\rm{RV}_{ H\alpha}$ and $\rm{RV}_{ H\beta}$  to the scale of $\rm{RV}_{ H\gamma}$ by subtracting 14.59 and 5.83 \kms from the measured values, respectively.

The final value of $\rm{RV}$ of a target is calculated by combining results from all  three Balmer lines, namely, 

\begin{equation}
\rm{RV}=\frac {\rm{RV}_{ H \alpha}/ \sigma_{\alpha}^2+\rm{RV}_{ H \beta}/ \sigma_{\beta}^2+\rm{RV}_{ H \gamma}/ \sigma_{\gamma}^2} {1 / \sigma_{\alpha}^2+1 / \sigma_{\beta}^2+1 / \sigma_{\gamma}^2}\text{,}
\end{equation}
\begin{equation}
\sigma_{\rm{RV}}=\sqrt {\frac {1} { 1 / \sigma_{\alpha}^2+1 / \sigma_{\beta}^2+1 / \sigma_{\gamma}^2}}\text{.}
\end{equation}

\begin{table*}
 \begin{minipage}{160mm}
 \centering
 \caption{Internal errors from three Balmer lines for different numbers for available measurements used in the fit}\label{tab:error}
\begin{footnotesize}
\begin{tabular}{cccc}
\hline
\hline
     $n$ & $\sigma_{\rm H\alpha}$ (\kms) & $\sigma_{\rm H\beta}$ (\kms)& $\sigma_{\rm H\gamma}$ (\kms) \\
\hline
       1              &       25.19&     15.81&    28.81\\
       2              &      20.17&       8.57&    20.46\\
       3              &      15.25&       6.57&    13.85\\
        $\geq 4$ &     12.05 &       5.72&    10.83\\
      \hline
\end{tabular}
\end{footnotesize}
\end{minipage}
\end{table*}

 %

 We note that the systemic radial velocities of  Type {\it c} RRLs can also derived, using the  template radial velocity curves for Types $ab$ RRLs, but only the systemic radial velocities of Type ab RRLs are recommended.

 \subsection{Validation with reference stars}
 
\begin{figure*}
  \centering
  \includegraphics[scale=0.45, angle=0]{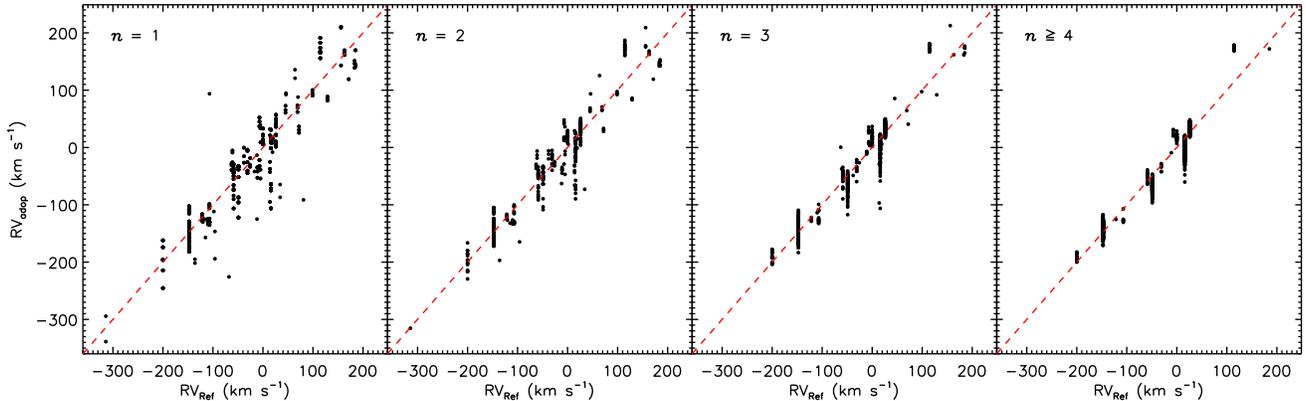}
  \caption{ Comparisons  of systemic radial velocities derived in the current work and those from literature.  
 The standard deviations are  20.6, 14.6, 9.7 and 4.5 \kms for $n$ = 1, 2, 3, and $\geq$ 4, respectively.
            }
   \label{fig:external}
\end{figure*}

 In order to check the uncertainties (both random and systemic) of our measurements, we collect RRLs from the literature with reliable radial velocity measurements 
 from high-resolution spectroscopy, GC members are also included
 \citep{Dambis2009, Harris2010, Kinman2012, Britavskiy2018}. In total, 108 common stars are found and their relevant information is listed in Table.\,\ref{tab:rvref} 
 (The information of 29 GC members has been listed in Tables\,\ref{tab:ref_star1}).
 Fig.\,\ref{fig:external}  shows the comparisons. The standard 
   derivations of the differences are  20.6, 14.6, 9.7 and 4.5 \kms for $n$ = 1, 2, 3, and $\ge$4, respectively. The median values of the differences are all around 2\,km\,s$^{-1}$, indicating no significant systemics of our final derived radial velocities.
We do not correct such a small offset since it is much smaller than the standard deviations.
   The results show that once the number of radial velocity measurements available to fit the radial velocity curve is great than 2, the standard deviation of the systemic velocity derived is likely to be less than 10 \kms.
  The expected uncertainty drops to  only 4.5 \kms for $n\geq 4$. This indicate our velocity measurements are quite robust.
  
 Although the pulsation nature of RRLs makes the determinations of their systemic radial velocities more difficult
than for other nonpulsating normal stars \citep[e.g.][]{Layden1994, Vivas2005, Prior2009}, the current work shows that one can still use the large numbers of low resolution spectroscopic observations to derive systemic radial velocities with a precisions between  5 and 21 \kms . 
This precision is adequate for  studying the Galactic halo properties considering that the line-of-sight velocity dispersion of
 halo stars is around 100 km s$^{-1}$ \citep[e.g.][]{Huang2016}.
  
  \begin{table*}
 \begin{minipage}{160mm}
 \centering
\caption{Radial velocities of reference stars selected from the literature} \label{tab:rvref} 
\begin{footnotesize}
\begin{tabular}{lccccc||lccccc}
\hline
\hline
 Star&RA & Dec & $\rm{RV}_{Ref}$ & error of $\rm{RV}_{Ref}$ & Reference & Star& RA & Dec  & $\rm{RV}_{Ref}$ & error of $\rm{RV}_{Ref}$ & Reference \\
        &(degree) & (degree) & (\kms) &    (\kms)       &  --   &  &(degree) & (degree) & (\kms) &    (\kms)       &  --  \\             
\hline
  DR Vir&16.295&  34.218&  $-$110.000&  4.000&      B18&
 XX And&19.364&  38.951&  0.000&  1.000&     D09\\
 CI And&28.785&  43.767&  24.000&  5.000&     D09&
 BK Eri&42.483&  $-$1.420&  141.000&  10.000&     D09\\
SS Tau&54.174&  5.361&  $-$11.000&  10.000&     D09&
 TU Per&47.270&  53.193&  $-$314.200&  0.500&     D09\\
AR Per&64.322&  47.400&  5.000&  1.000&     D09&
SZ Vir&118.431&  19.273&  346.000&  6.000&      B18\\
RR Gem&110.390&  30.883&  64.000&  1.000&     D09&
YY Lyn&116.375&  37.383& $-$93.100&  15.000&     K12\\
ZZ Lyn&117.591&  37.700&  147.000&  15.000&     D09&
VY Lyn&113.108&  38.835&  114.500&  15.000&     K12\\
AC Lyn&118.675&  38.906&  $-$25.800&  15.000&     K12&
VX Lyn&112.966&  39.130&  1.000&  15.000&     D09\\
WX Lyn&113.910&  39.257&  26.000&  15.000&     D09&
TZ Aur&107.896&  40.777&  45.000&  2.000&     D09\\
VZ Lyn&113.170&  41.627&  $-$182.100&  15.000&     K12&
TW Lyn&116.276&  43.112&  $-$39.000&  1.000&     D09\\
DD Hya&123.132&  2.835&  156.000&  1.000&     D09&
AN Cnc&134.543&  15.805&  16.000&  14.000&     D09\\
AF Lyn&128.989&  41.020&  $-$121.700&  15.000&     K12&
P 54-13&120.484&  41.022&  69.000&  10.000&    K12\\
T Sex&148.369&  2.057&  29.000&  1.000&      D09&
BS 16927&146.151&  41.143&  70.000&  10.000&  K12\\
TT Lyn&135.784&  44.586&  $-$65.000&  2.000&     D09&
SW Leo&163.981& $-$2.982&  46.000&  11.000&     D09\\
BK UMa&162.579&  42.569&  171.400&  5.000&     K12&
TV Leo&167.841&  $-$5.892& $-$96.000&  5.000&     D09\\
SS Leo&173.477& $-$0.033&  163.000&  2.000&     D09&
SZ Leo&165.404&  8.166&  185.000&  4.000&     D09\\
AE Leo&171.551&  17.661&  $-$53.000&  10.000&     D09&
BN UMa&169.095&  41.234&  19.000&  10.000&     K12\\
BQ Vir&189.114&  $-$2.426&  129.000&  9.000&     D09&
UV Vir&185.320&  0.368&  99.000&  11.000&     D09\\
FU Vir&189.610&  13.016&  $-$90.000&  8.000&     D09&
S Com&188.190&  27.029& $-$55.000&  1.000&      D09\\
DV Com&190.977&  28.021&  $-$136.000&  10.000&     D09&
EO Com&194.342&  28.889&  72.000&  10.000&     D09\\
EM Com&192.910&  30.518&  $-$127.000&  10.000&     D09&
CD Com&183.142&  30.801&  $-$203.000&  10.000&     D09\\
TU Com&183.446&  30.985&  $-$98.000&  10.000&     D09&
CK UMa&180.402&  31.903&  16.000&  10.000&     K12\\
CK Com&183.711&  33.102&  $-$88.000&  10.000&     D09&
DC CVn&191.818&  35.202& $-$200.000&  10.000&     D09\\
SW Vir&190.229&  37.085&  14.000&  6.000&      B18&
UZ Vir&187.615&  40.509&  $-$49.000&  6.000&      B18\\
Z CVn&192.439&  43.774&  14.000&  10.000&      D09&
WW Vir&202.099&  $-$5.286&  129.000&  10.000&     D09\\
BC Vir&200.588&  5.886&  4.000&  13.000&     D09&
BB Vir&207.920&  6.431&  $-$38.000&  13.000&     D09\\
AV Vir&200.048&  9.188&  153.000&  1.000&     D09&
UY Boo&209.693&  12.952&  145.000&  2.000&     D09\\
ST Com&199.464&  20.781&  $-$68.000&  10.000&     D09&
RY Com&196.283&  23.278& $-$31.000&  8.000&     D09\\
ST CVn&209.392&  29.858&  $-$129.000&  1.000&     D09&
EW Com&198.257&  31.023&  19.000&  10.000&     D09\\
RZ CVn&206.263&  32.655&  $-$12.000&  1.000&     D09&
SS CVn&207.066&  39.901&  $-$40.000&  3.000&     D09\\
RV Vir&203.325&  53.988&  $-$175.000&  6.000&      B18&
SX UMa&201.556&  56.257& $-$154.000&  1.000&     D09\\
AE Vir&216.872&  3.778&  208.000&  10.000&     D09&
RS Boo&218.389&  31.755&  $-$7.000&  1.000&     D09\\
TW Vir&221.275&  41.029&  $-$89.000&  2.000&      B18&
AN Ser&238.379&  12.961&  81.000&  1.000&           C17\\
BH Ser&228.754&  19.443&  $-$113.000&  11.000&     D09&
AW Ser&241.620&  15.368&  $-$126.000&  15.000&     D09\\
RW Dra&248.882&  57.840& $-$112.000&  1.000&     D09&
V0816 Oph&265.658&  4.958& $-$28.000&  10.000&  D09\\
V0784 Oph&263.855&  7.756&  $-$167.000&  10.000&  D09&
DL Her&260.094&  14.511&  $-$61.000&  14.000&     D09\\
TW Her&268.630&  30.410&  $-$5.000&  1.000&     D09&
VZ Her&258.267&  35.979&  $-$115.000&  1.000&     D09\\
KX Lyr&278.314&  40.173&  $-$36.200&  0.500&     D09&
SX Aqr&324.035&  3.231&  $-$165.000&  3.000&     D09\\
AO Peg&321.765&  18.599&  264.000&  4.000&     D09&
VV Peg&333.266&  18.451&  34.200&  0.000&     D09\\
DZ Peg&350.029&  16.069& $-$289.500&  2.400&     D09&
VZ Peg&355.568&  24.916&  $-$264.000&  1.000&     D09\\
BK And&353.775&  41.103&  $-$17.000&  7.000&     D09&
           &              &             &              &           &            \\                      
\hline \hline
\end{tabular}
\begin{flushleft}
\textbf{Notes.} The references are: B18-- \citet{Britavskiy2018}; D09--\citet{Dambis2009}; K12-- \citet{Kinman2012}; C17--\citet{Chadid2017}.
\end{flushleft}
\end{footnotesize}
\end{minipage}
\end{table*}

\subsection{Discussion}
 \citet{Sesar2012} shows the amplitude of the template radial velocity curve of RRLs is a function of the amplitude of ${\rm V}$-band light curve ( A$_V$) . However, only part of the RRLs in our catalog with A$_V$ available, we therefore fixed the amplitude of the template radial velocity curve in the fitting to derive the systemic radial velocity (see Section\,4.1). To evaluate the effects of our constant radial velocity curve amplitude choice on deriving the systemic radial velocities of RRLs, we re-determine the systemic radial velocities of $>2000$ RRLs with A$_V$ known. At this time, we repeat the fitting described in Section\,4.1 by using the values of ${\rm A_{rv}}$ determined by A$_V$ from the functions provided by \citet{Sesar2012} and obtain their new systemic radial velocities.
 
By comparing the new results with our adopted ones, the median velocity difference is found around 2\,km\,s$^{-1}$ and the standard deviation of the velocity difference is only about 4\,km\,s$^{-1}$.
According to the above test, our constant radial velocity curve amplitude choice show minor effects on deriving the systemic radial velocities of RRLs.
 
\textbf{The new values of systemic radial velocity determined in this section is also released in the final catalog for reference.  }

\section{Distances}
\subsection {Measurement method}
 Generally, RRLs are divided into three types: Type ab RRLs (RRab) pulsates in the fundamental mode and 
 Type c RRLs (RRc) pulsates in first-overtone mode and Type d RRLs (RRd) pulsates in both modes simultaneously.
In our sample, the number of RRd is small, so we do not discuss them separately.
 Being standard candles, RRLs obey the well-defined relations  between 
 absolute visual magnitude and metallicity,

 \begin{equation}
M_V= a\, \rm{[Fe/H]} + \textit{b} \text{,}
\label{eq:Mvab}
\end{equation}
and the near- or mid-infrared period-absolute magnitude-metallicity ($PMZ$),
 
 \begin{equation}
M_{K_s/W_1}=c\, \rm{log(P)}+\textit{d}\,  \rm{[Fe/H]}+ \textit{e} \text{.}
\end{equation}

  Coefficients $a, b, c, d, e$ are different for Type ab and Type c RRLs.
  In this paper, for RRab, coefficients $a, b, c, d, e$ are taken from \citet{Muraveva2018}, derived using the latest data of Gaia DR2 (please refer to their Table 4 for more detail). 
  For RRc, coefficients $a, b$ given by \citet{Ferro2017} are used. We adopt the mid-infrared period-absolute magnitude relation provided by  \citet{Klein2014},
  
  \begin{equation}
M_{W1}=-1.64\,log(\rm{P}/0.32)-\,0.231\text{.}
\label{eq:Mw1c}
\end{equation}

Compared to the visual $M_V$ -  \rm{[Fe/H]} relation, the near- or mid-infrared $PMZ$ relations
 is less affected by the interstellar extinction. In the following calculations, we prefer distances derived from the near/mid-infrared $PMZ$ relations as the final ones. 

We cross match our sample with the 2MASS and WISE catalogs to obtain apparent magnitudes in $K_{\rm s}$ and $W1$ bands for our sample stars. For sources selected from the Catalina survey, 
 Catalina $V$ band magnitudes are converted to Johnson $V$ band magnitudes using the transformation equation of \citet{Graham2015}, assuming an intrinsic $(B-V)_0$ color of 0.2 for RRLs.
 We then use above relations to obtain the absolute magnitudes and further derive the distances, after the interstellar extinction corrections. 
 
 Actually, the WISE magnitudes could been taken as mean magnitudes since they are the mean values typically of $\geq$ 10 single-epoch observations obtained by the WISE survey \citep{Wright2010}. We therefore derive the distances of RRLs from $W1$ band if uncertainties smaller than 0.1 mag. For only 13 stars without good $W1$ but $K_{\rm s}$ band photometry, we derive their distances from $K_{\rm s}$ band. If both $K_{\rm s}$ and $W1$ band photometric uncertainties do not meet our requirements, we use the visual $M_V$ -  \rm{[Fe/H]} relation to obtain the distance.


Totally, we obtain distances for 4,919 RRLs with the above procedures ( 4,061 from $W1$ band, 13 from $K_{\rm s}$ band and 845 from $V$ band).

\subsection {Validation of distances}
\subsubsection{Comparison with reference stars}

\begin{figure*}
  \centering
  \includegraphics[width=16.0cm, angle=0]{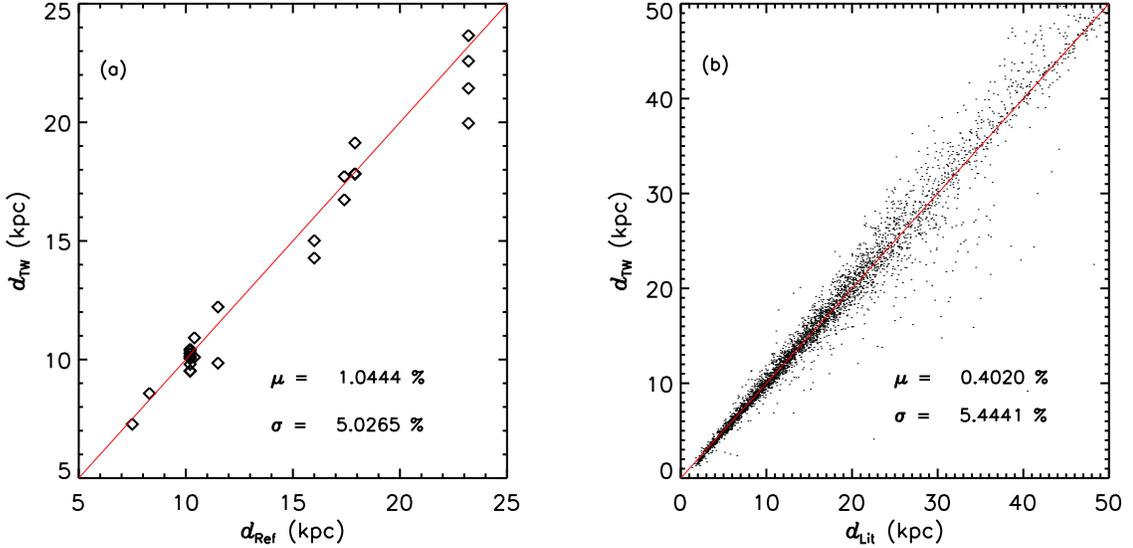}
 \caption { Panel (a) compares distances derived in this work ($d_{\rm{TW}}$) with reference values from the star cluster catalog ($d_{\rm{Ref}}$). Panel (b) comparison but with reference values collected from other sources in the literatures ($d_{\rm{Lit}}$).
              }

   \label{fig:distance}
\end{figure*}

 In order to check the reliability of our calculations, we firstly compare the distances obtained above with those in the literature for the reference sample,
 for which the reference distances obtained from the star cluster catalog.
  As shown in Fig.\,\ref{fig:distance}, the average relatives difference ($\frac {\Delta d} {d}$) is about 1 per cent, with a dispersion of about 5 per cent.
  We also collect information of distances from other sources in the literatures \citep{Vivas2006, Miceli2008, Watkins2009, Suveges2012, Drake2013, Drake2013b,Sesar2013}. The comparison is also shown in Fig. \ref {fig:distance}.
  We find that they all agree very well with each other.
   
\subsubsection{Comparison with Gaia parallaxes}

\begin{figure*}
  \centering
  \includegraphics[width=16.0cm, angle=0]{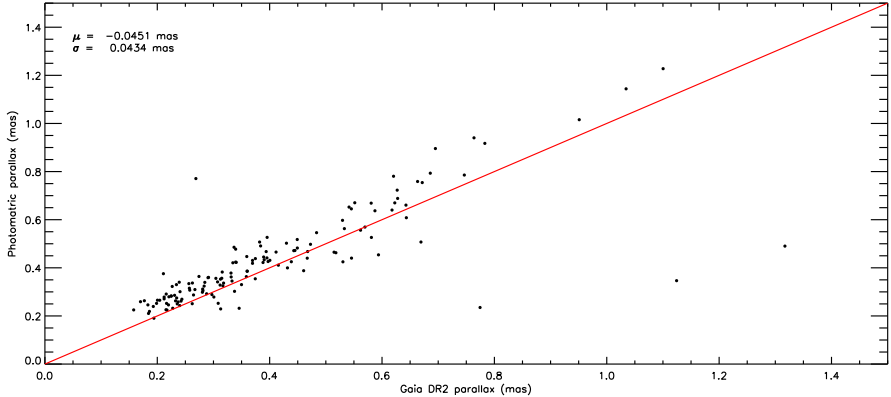}
  \caption{ Comparison of parallaxes converted from distances derived in the current work and those measured by the Gaia DR2 for RRab stars.} 
   \label{dRRab}
\end{figure*}

 We also compare our distances measurements with the Gaia parallaxes. We first convert our distances estimates of uncertainties less than 10 per cent, derived from near- or mid infrared or visual magnitudes, into photometric parallaxes, then compare them with those from the Gaia DR2. 
 Fig. \ref{dRRab} shows the comparison for RRab stars. The mean deviation is $-0.04$ mas, consistent with the offset of zero point reported by the Gaia collaboration \citep{Gaia2018}. As seen in Fig.\,\ref{dRRab}, four outliers are significantly deviate the one-to-one line. We have checked the distances derived here to those from the literatures and find they are all in great consistent. The reason of those outliers are still unclear and one possible explanation is that their parallax solutions are somehow wrong. Similar results are found for RRc stars.
 
 \section{The final catalog}
\begin{figure*}
  \centering
  \includegraphics[width=16.0cm, angle=0]{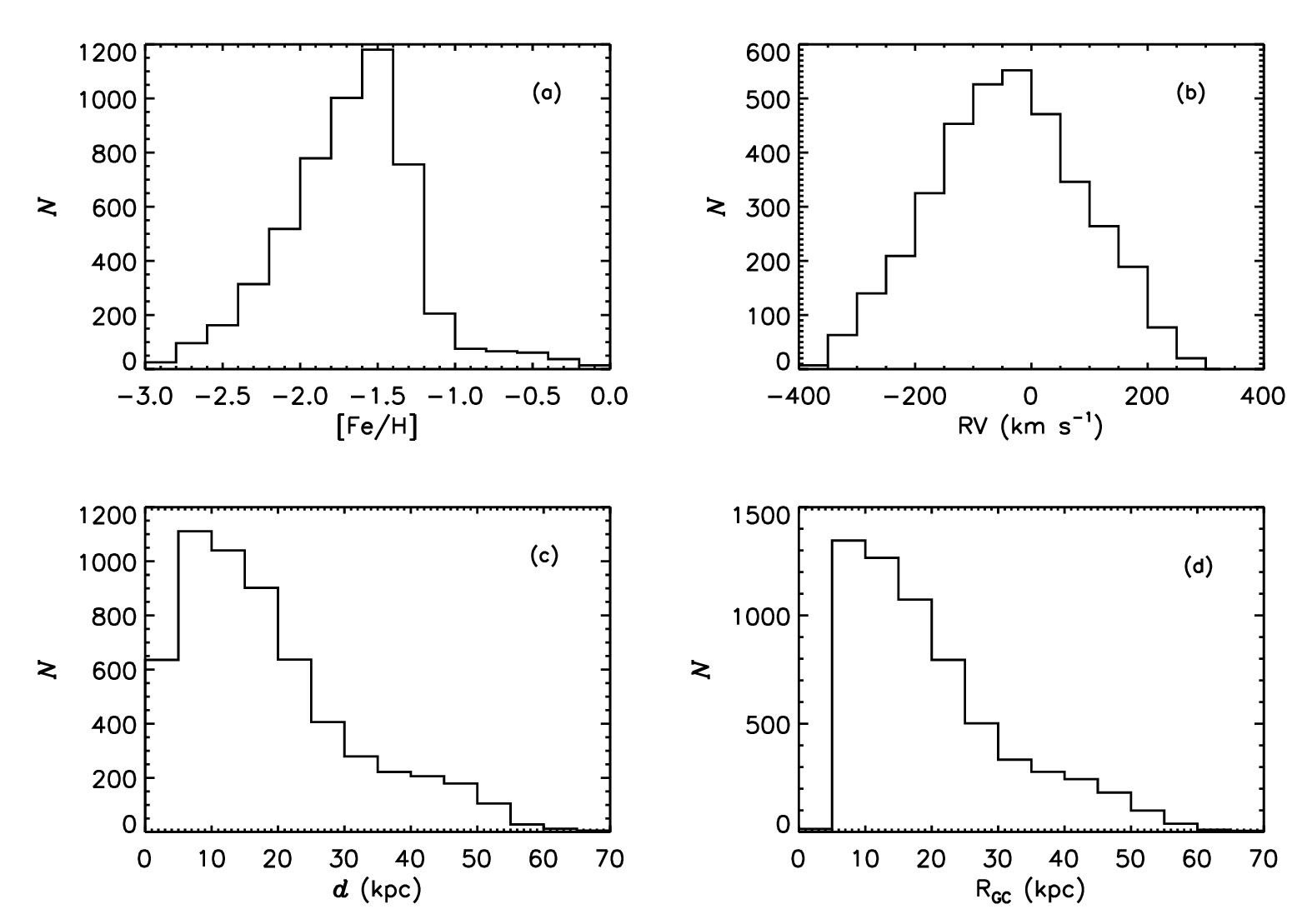}
  \caption{  Panel shows the distributions of metallicities (a), systemic radial velocities (b),
                  distance (c) and distances to the center of the Galaxy (d) derived in the current work for stars in the final, main catalog.} 
   \label{fig:final}
\end{figure*}

  By combing the SDSS and LAMOST spectroscopic data  with literature the photometric data of objects, 
  we have compiled a catalog of RRLs containing 6,268 uniq RRLs.  For objects with individual single-exposure SNRs greater than 10 and not affected by 
  shock waves, we measure the metallicities by a least $\chi^2$ fitting technique. The weighted mean metallicity of the individual spectra, are adopted as the final values. 
  In total, metallicities for 5,290 RRLs are obtained in this way. 
   We then fit the  templates of systemic velocity of RRLs provided by \citet{Sesar2012} to the observed ones and derive 
  the systemic radial velocity for 3,642 RRLs. Finally, we use the $PMZ$ or $M_{\rm V}$- [Fe/H]  relations  to calculate distances of our sample stars
  and obtain distance estimates for 4,919 RRLs.
  All the information derived is compiled into an online catalog containing 6,268 RRLs (Table 6). Table \ref{tab:final} lists the columns contained in the main catalog.
   Fig. \ref{fig:final} shows the distributions of metallicities, systemic radial velocities, distances and  Galactocentric distances of stars in the final, main catalog. 
   
   In the near future, we plan to enlarge our sample by including data from additional spectroscopic surveys, e.g., the GALAH survey  \citep{DeSilva2015}, as well as additional photometrically identified RRLs from the recent time-domain surveys, e.g. the Pan-STARRS1 \citep{Chamber2016, Sesar2017}; the Gaia DR2 \citep{Clementini2019}.
   We expect to have precise measurements of metallicities, systemic radial velocities and distances of RRLs for the whole sky. The data shall be very helpful for the  study of the formation and evolution of the Galactic halo.

\begin{table*}
\begin{threeparttable}
\caption{The columns of RRLs catalog} \label{tab:final}
\centering 
\begin{tabular}{l l l l} 
\hline
\hline
&Column&Unit&Description\\
\hline
1&ID& -&A unique object id for the cataloged RRLs\\
2&RA&degree&Right ascension at J2000 from the photometric surveys\\
3&DEC&degree&Declination at J2000 from the photometric surveys\\
4&GL&degree&Galactic longitude \\
5&GB&degree&Galactic latitude \\
6&VMAG&mag&$V$-band magnitude\\
7&ERR-VMAG&mag&Error in $V$-band magnitude\\
8&PER& day& Period\\
9&PFROM& -& Reference for period \\
10&EPOCH&day &Date of maximum light from the photometric surveys\\
11&EPOCH-FLG&-& Which type of EPOCH (MJD,JD, HJD)\\
12&VARTYPE&-& Type\\
13&SNR& -& Spectral SNR at 4650 \AA\\
14&FEH-ADOP& dex& Metallicity derived in the current work\\
15&ERR-FEH-ADOP&dex& Error of FEH-ADOP\\
16&FEH-REF&dex& Metallicity from the literature if available\\
17&ERR-FEH-REF&dex & Error of FEH-REF\\
18&RV-ADOP \tnote{1}& \kms& Systemic radial velocity derived in the current work\\
19& ERR-RV-ADOP&\kms& Error of RV-ADOP \\
20&RV-REF1& \kms& Systemic radial velocity determined in Section 4.3\\ 
21&ERR-RV-REF1&\kms& Error of RV-REF1\\
22&RV-REF2& \kms& Systemic radial velocity from the literature if available\\
23&ERR-RV-REF2&\kms& Error of RV-REF2\\
24&DIST-ADOP& kpc& Distance derived in the current work\\
25&ERR-DIST-ADOP& kpc& Error of DIST-ADOP\\
26 &DIST-REF& kpc& Distance from the literature if available\\
27&ERR-DIST-ADOP& kpc& Error of DIST-ADOP\\
28&PMRA                 &  $\rm{mas}\,yr^{-1}$ & Proper motion in $\alpha\cos\delta$ from Gaia DR2\\
29&PMDEC              & $\rm{mas}\,yr^{-1}$ & Proper motion in $\delta$ from Gaia DR2\\
30& ERR-PMRA      & $\rm{mas}\,yr^{-1}$ & Error of the proper motion in $\alpha\cos\delta$ from Gaia DR2\\
31&ERR-PMDEC     & $\rm{mas}\,yr^{-1}$ & Error of the proper motion in $\delta$ from Gaia Dr2\\
32&NUMBER           &--                               & Number of individual spectra available\\
\hline
\hline
\end{tabular}
\begin{tablenotes}
   \item[1] The radial velocities for type c RRLs are not recommended for use.
 \end{tablenotes}
 \end{threeparttable}
\end{table*}

\section{Conclusion}

We present a catalog of 5,290 RRLs with metallicity estimates. Nearly 70 per cent of them also have systemic radial velocity measurements. 
We  use the single-exposure spectra rather than the combined spectra for metallicity and velocity estimation.
 We develop a criterion based on the measured equivalent width of Ca~{\sc ii} K line, EW(Ca~{\sc ii} K), to diagnose whether
 a spectrum is affected by shock waves or not. Those affected are excluded, from the metallicity and velocity estimation. 
  
 We measure the systemic radial velocities using the empirical template radial velocity curves of RRLs provided by \citet{Sesar2012} and obtain results for 3,642 RRLs in total. The typical error is about $5 \sim 21$ \kms, dependent on the number of radial velocity measurements available at different phases. 
  
 Finally, with the well calibrated near-infrared $PM_{K_{\rm s}}Z$ or $PM_{W1}Z$, and $M_{V}$-[Fe/H] relations, precise distances are derived for 4,919 RRLs.
 
 The results provide vital information to study many issues related to the Galactic halo.

 \acknowledgements
We thank the anonymous referee for his/her helpful suggestions and comments.
This work is supported by the National Science Foundation of China under grant No. U1731108,
and Natural Science Foundation of CTGU under grant No. KJ2014B078.
This work is also supported by National Key R\&D Program of China No. 2019YFA0405500.
 Y. H., B.Q.C. and X.W.L. acknowledge the National Natural Science Foundation of China U1531244, 11833006, 11811530289 and 11903027.
 H.W.Z. acknowledge the National Natural Science Foundation of China 11973001.
The Guoshoujing Telescope (the Large Sky Area Multi-Object Fiber
Spectroscopic Telescope LAMOST) is a National Major Scientific
Project built by the Chinese Academy of Sciences. Funding for
the project has been provided by the National Development and
Reform Commission. LAMOST is operated and managed by the National
Astronomical Observatories, Chinese Academy of Sciences. 

This work has made use of data from the European Space Agency (ESA) mission Gaia
 (https://www.cosmos.esa.int/gaia), processed by the
Gaia Data Processing and Analysis Consortium (DPAC,
https://www.cosmos.esa.int/web/gaia/dpac/consortium).
This work also has made use of data products from the SDSS,
2MASS, and WISE.

\bibliographystyle{apj}

\bibliographystyle{yahapj}

\end{document}